\documentclass[
    ,final            
  ]
  {aipproc}

\layoutstyle{6x9}

\begin{document}

\title{Modeling non-Gaussian 1/f Noise by the Stochastic Differential Equations}


\author{B. Kaulakys}{
  address={Institute of Theoretical Physics and Astronomy of Vilnius University, A. Gostauto 12, LT-01108 Vilnius, Lithuania}
}

\author{M. Alaburda}{
  address={Institute of Theoretical Physics and Astronomy of Vilnius University, A. Gostauto 12, LT-01108 Vilnius, Lithuania}
}

\author{J. Ruseckas}{
  address={Institute of Theoretical Physics and Astronomy of Vilnius University, A. Gostauto 12, LT-01108 Vilnius, Lithuania}
}

\begin{abstract}
We consider stochastic model based on the linear stochastic differential equation with the linear relaxation and with the diffusion-like fluctuations of the relaxation rate. The model generates monofractal signals with the non-Gaussian power-law distributions and $1/f^{\beta}$ noise.
\end{abstract}

\maketitle


\section{Introduction}

The presence of $1/f$ noise is ubiquitous in a variety of different systems. Mostly $1/f$ noise is Gaussian, but sometimes the signals exhibiting $1/f$ fluctuations are non-Gaussian. The non-Gaussianity is often taken as a signature of fluctuator's interaction \cite{Parman1992}. Nevertheless, statistically independent and noninteracting fluctuators may exhibit non-Gaussian noise, as well \cite{Seidler1996}, especially when the fluctuations are strong \cite{Kaulakys2005,Orlyanchik2006}. 

Recently we proposed stochastic models of $1/f$ noise based on the nonlinear stochastic differential equations \cite{Kaulakys2004,Kaulakys2006}. The models generate signals with the power-law distributions of the signal intensity and the power-law spectral densities.

Moreover, $1/f$ noise is often represented as a sum of independent Lorentzian spectra resulting from uncorrelated components of the signal with a wide-range distribution of the relaxation times \cite{Bernamont1937}. It should be noted that summation of the spectra is allowed only if the processes with different relaxation times are isolated from each another \cite{Hooge1997}. Distribution densities of the signal components described by the linear stochastic differential equation are Gaussian and the distribution density of the signal resulting from the similarly distributed components is usually Gaussian as well \cite{Kaulakys2005}. However, the signal consisting of the sequence of components with very different variances may be non-Gaussian. In this paper we will analyze the non-Gaussianity of the signals exhibiting $1/f$ noise and generated by the linear stochastic differential equations with the fluctuating relaxation rate.

\section{The model}

Consider the random process $x$ described by the stochastic differential equation
\begin{equation}
dx=-\gamma(t)xdt+\sigma dW \label{diff_eq}
\end{equation}
with the time dependent relaxation rate $\gamma(t)$. Here $W$ is the Wiener process, $dW=\xi(t)dt$, with $\xi(t)$ being the $\delta$-correlated white noise, $\langle\xi(t)\xi(t^{\prime})\rangle=\delta(t-t^{\prime})$, and $\sigma$ is the intensity (standard deviation) of the white noise. In this paper the stochastic differential equations we understand in Ito interpretation. 

When the relaxation rate changes very slowly, we have the signal as a sequence of signals with different relaxation rates.

We can eliminate the parameter $\sigma$ by the appropriate change of the time scale, $t\to\sigma^{2}t$, while the change of the relaxation rate $\gamma(t)$ may be described by another stochastic differential equation, resulting, e.g., in the power-law distribution of $\gamma$. 

Therefore, we have the system of two equations,
\begin{equation}
d\gamma=\sigma_{\gamma}\gamma^{\mu}dW_{\gamma}, \label{diff_gamma}
\end{equation}
\begin{equation}
dx=-\gamma xdt+dW. \label{diff_x}
\end{equation}

Here $\sigma_{\gamma}$ determines the speed of the change of the relaxation rate $\gamma$ driven by the white noise $\xi_{\gamma}$ and the factor $\gamma^{\mu}$ imposes the power-law distribution, $P_{r}(\gamma)\sim\gamma^{\eta}$ (with $\eta=-2\mu$), of the relaxation rate. The diffusion-like motion of $\gamma$ should be restricted in some interval, e.g., (0,1). Then
\begin{equation}
P_{r}(\gamma)=(1+\eta)\gamma^{\eta}. \label{dist}
\end{equation}

We can restrict the analysis of the positive $x$, only, with the reflection of $x$ at $x=0$. For the definite $\gamma$ the distribution of $x$ is Gaussian and the power spectrum is Lorentzian,
\begin{equation}
P_{1}(\gamma|x)=2\sqrt{\frac{\gamma}{\pi}}e^{-\gamma x^{2}}, \label{dist_1}
\end{equation}
\begin{equation}
S_{1}(\gamma|f)\sim(\gamma^{2}+\omega^{2})^{-1}. \label{spec_1}
\end{equation}
For very slow evolution of the relaxation rate $\gamma$, the resulting characteristics of the system \eqref{diff_gamma} and \eqref{diff_x} yields from the average of expressions \eqref{dist_1} and \eqref{spec_1} over distribution \eqref{dist}, 
\begin{equation}
P(x)=\int P_{1}(\gamma|x)P_{r}(\gamma)d\gamma=\frac{2(1+\eta)}{\sqrt{\pi}x^{3+2\eta}}\left[ \Gamma \left( \frac
32+\eta \right) -\Gamma \left( \frac 32+\eta ,x^2\right) \right], \label{dist_gen}
\end{equation}
\begin{equation}
S(f)=\int P_{1}(\gamma|f)P_{r}(\gamma)d\gamma\sim 1/f^{1-\eta},\quad f\ll 1. \label{spec_gen}
\end{equation}
Here $\Gamma(a,x)$ is the incomplete gamma function.

Therefore, the simple linear stochastic equation \eqref{diff_x} with the additive noise, linear relaxation and the power-law distribution near zero of slowly changing relaxation rate results asymptotically in the power-law distribution of the signal,
\begin{equation}
P(x)\sim 1/x^{3+2\eta},\quad x\gg 1, \label{dist_asympt}
\end{equation}
and power-law distribution \eqref{spec_gen} of the low frequency spectrum.

For the uniform distribution of the relaxation rate $\gamma$, i.e., for $\mu=\eta=0$, we have from Eqs. \eqref{dist_gen} and \eqref{spec_gen}
\begin{equation}
P(x)=\frac{1}{x^{3}}\mathop{\mathrm{erf}} x-\frac{1}{x^{2}}\exp(-x^{2}), \label{dist_uniform}
\end{equation}
\begin{equation}
S(f)\sim\arctan(1/2\pi f)/f. \label{spec_uniform}
\end{equation}

\section{Numerical analysis}

\begin{figure}
  \hspace{-10pt}
  \includegraphics[width=.5\textwidth]{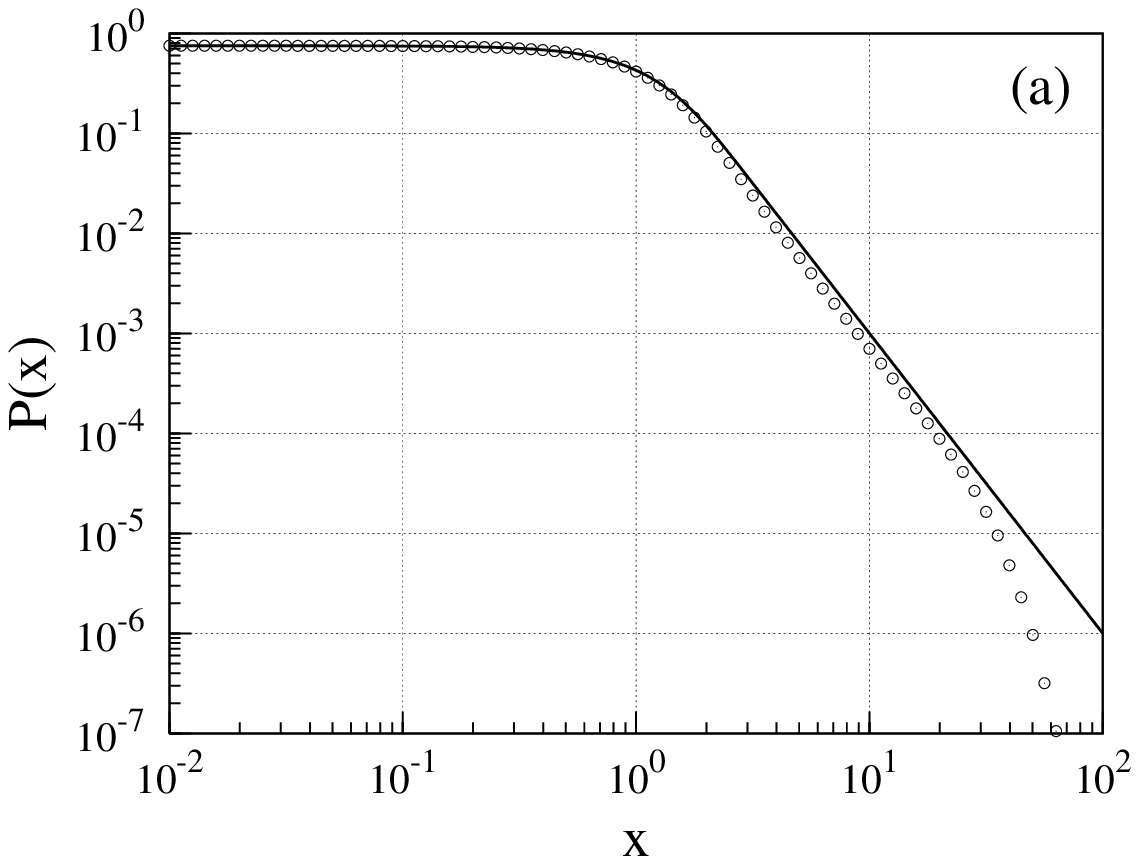}
  \hspace{-10pt}
  \includegraphics[width=.5\textwidth]{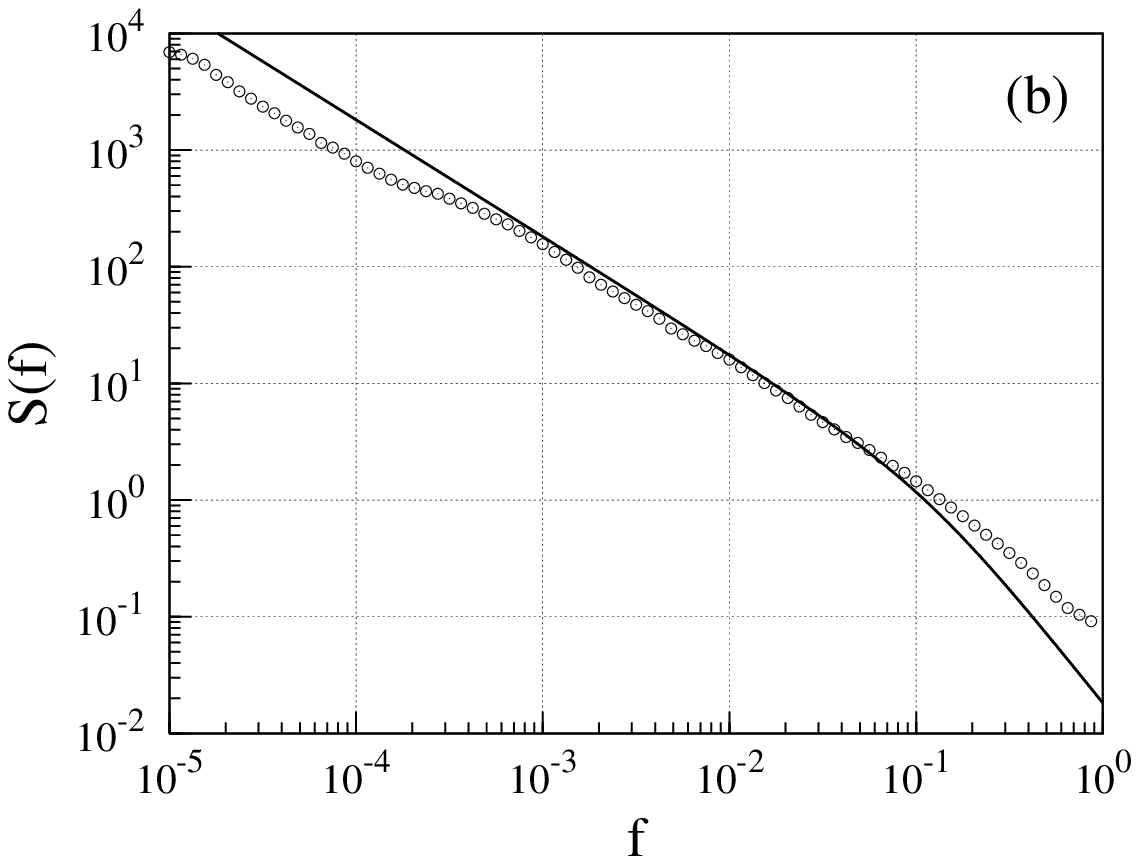}
  \caption{Probability density (a) and power spectrum (b) of the signal generated according to Eqs. \eqref{diff_gamma} and \eqref{diff_x} with $\mu=0$ and $\sigma_{\gamma}=2\cdot 10^{-4}$ in comparison with the analytical expressions \eqref{dist_uniform} and \eqref{spec_uniform}.}
\label{fig:mu0}
\end{figure}
\begin{figure}
  \hspace{-10pt}
  \includegraphics[width=.5\textwidth]{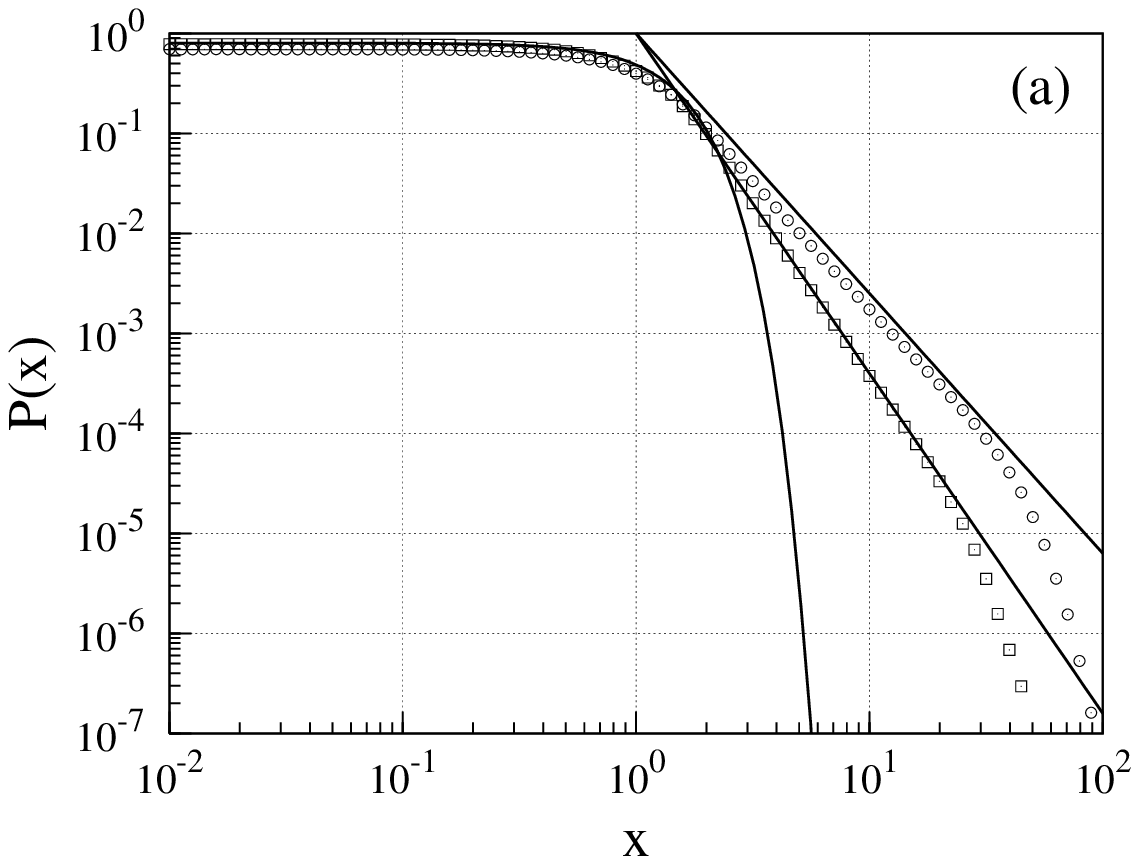}
  \hspace{-10pt}
  \includegraphics[width=.5\textwidth]{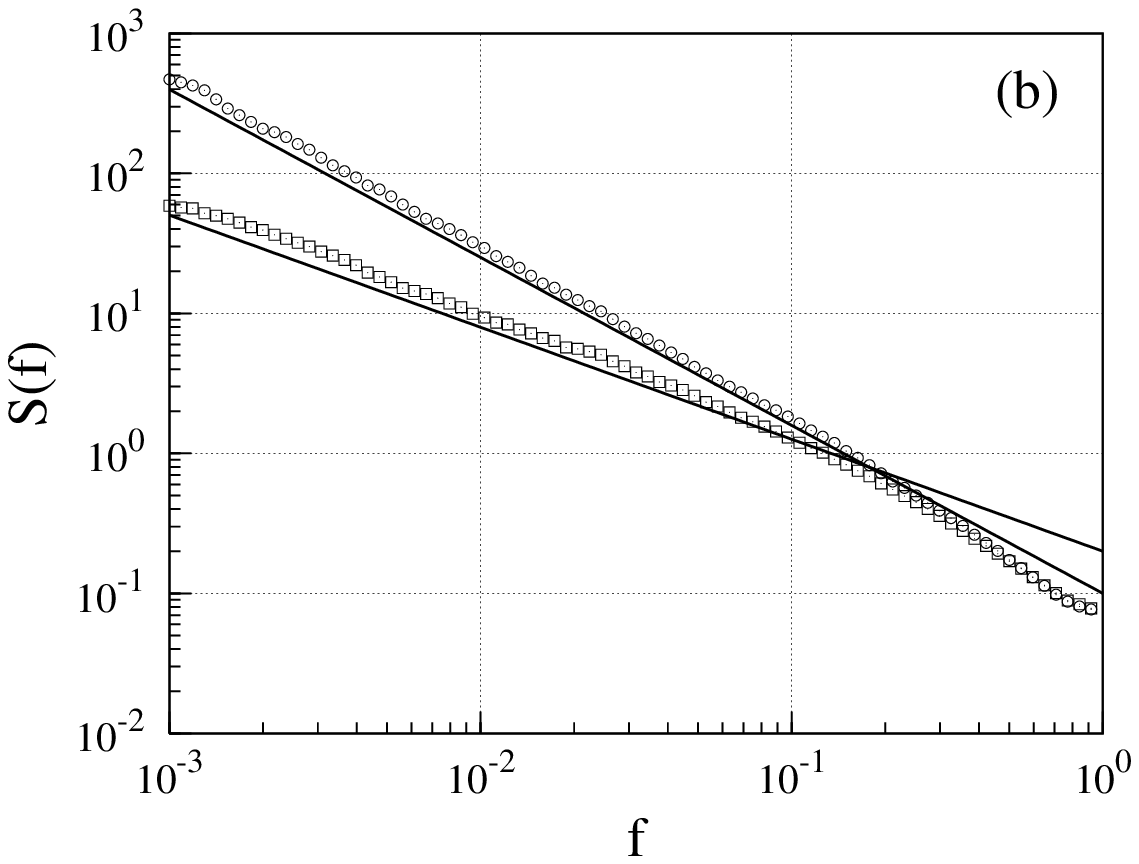}
  \caption{As in Fig. \ref{fig:mu0}, but with $\mu=0.1$, open circles, and $\mu=-0.1$, open squares, in comparison with Eqs. \eqref{spec_gen} and \eqref{dist_asympt}. The lowest solid curve represents Gaussian distribution.}
\label{fig:mu}
\end{figure}

We have performed numerical analysis of the model \eqref{diff_gamma} and \eqref{diff_x}, as well. In figure \ref{fig:mu0} the simulation results for $\mu=0$, i.e., for the case of pure $1/f$ noise are presented. We see good agreement with the analytical expressions in large intervals of frequency and distribution of the signal. Figure \ref{fig:mu} shows the dependences of distribution and the slope of the spectral density on the parameter $\mu$. In all cases the distribution density of the signal exhibits the ``fat tail'' distributions in contrast to the short-range Gaussian distribution for the fixed relaxation rate $\gamma$.

We analysed the multifractality of the signals, as well. For this purpose we calculated a generalized $q$th order height-height correlation function (GHCF) $F_{q}(t)$ defined as \cite{fractality}
\begin{equation}
F_{q}(t)=\langle|I(t^{\prime}+t)-I(t^{\prime})|^{q}\rangle^{1/q}, 
\label{ghcf}
\end{equation}
where the angular brackets denote the time average. The GHCF $F_{q}(t)$ characterizes the correlation properties of the 
signal $I(t)$, and for a multiaffine signal a power-law behavior, 
\begin{equation}
F_{q}(t)\sim t^{H_{q}},
\label{ghcf_approx}
\end{equation}
is expected. Here $H_{q}$ is the generalized $q$th order Hurst exponent. 
If $H_{q}$ is independent on $q$, a single scaling exponent $H_{q}$ is involved, and the signal $I(t)$ is said to be 
monofractal \cite{fractality,fractality1}. 
If $H_{q}$ depends on $q$, the signal is considered to be multifractal. 

\begin{figure}
  \hspace{-10pt}
  \includegraphics[width=.5\textwidth]{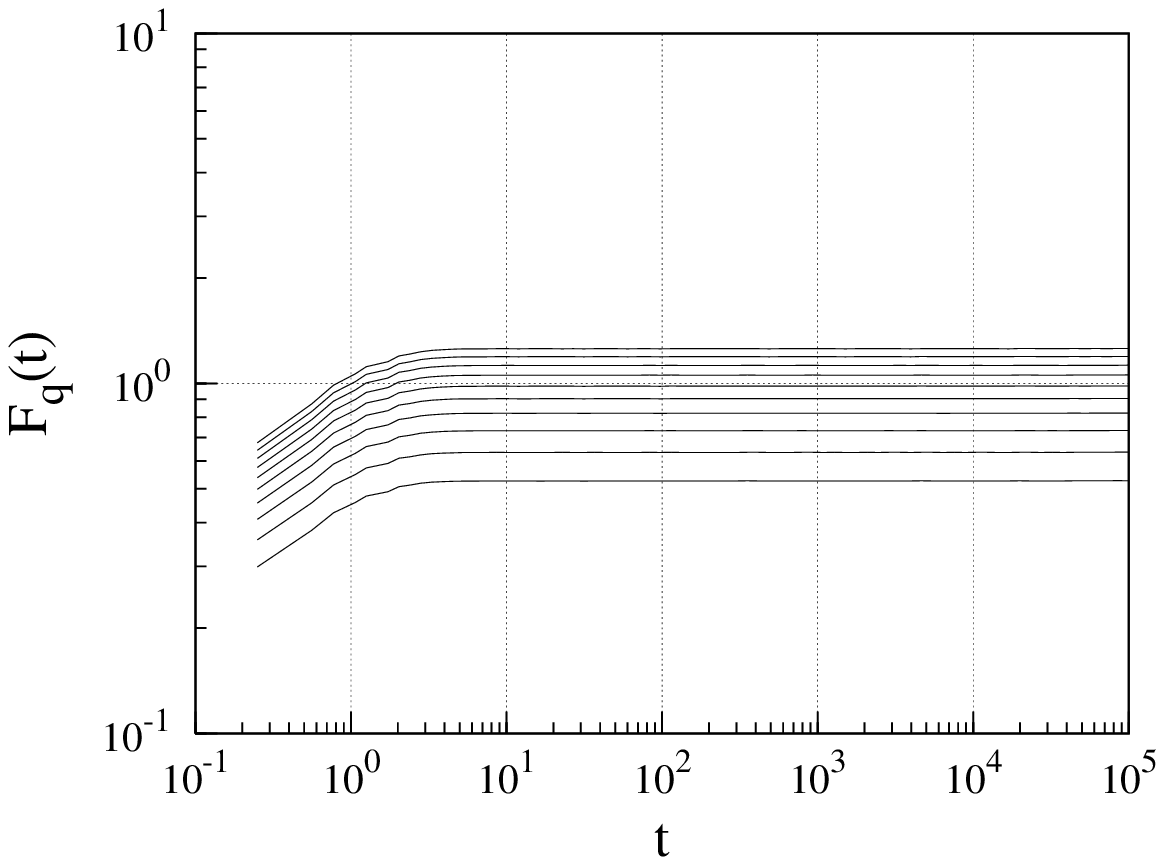}
  \hspace{-10pt}
  \includegraphics[width=.5\textwidth]{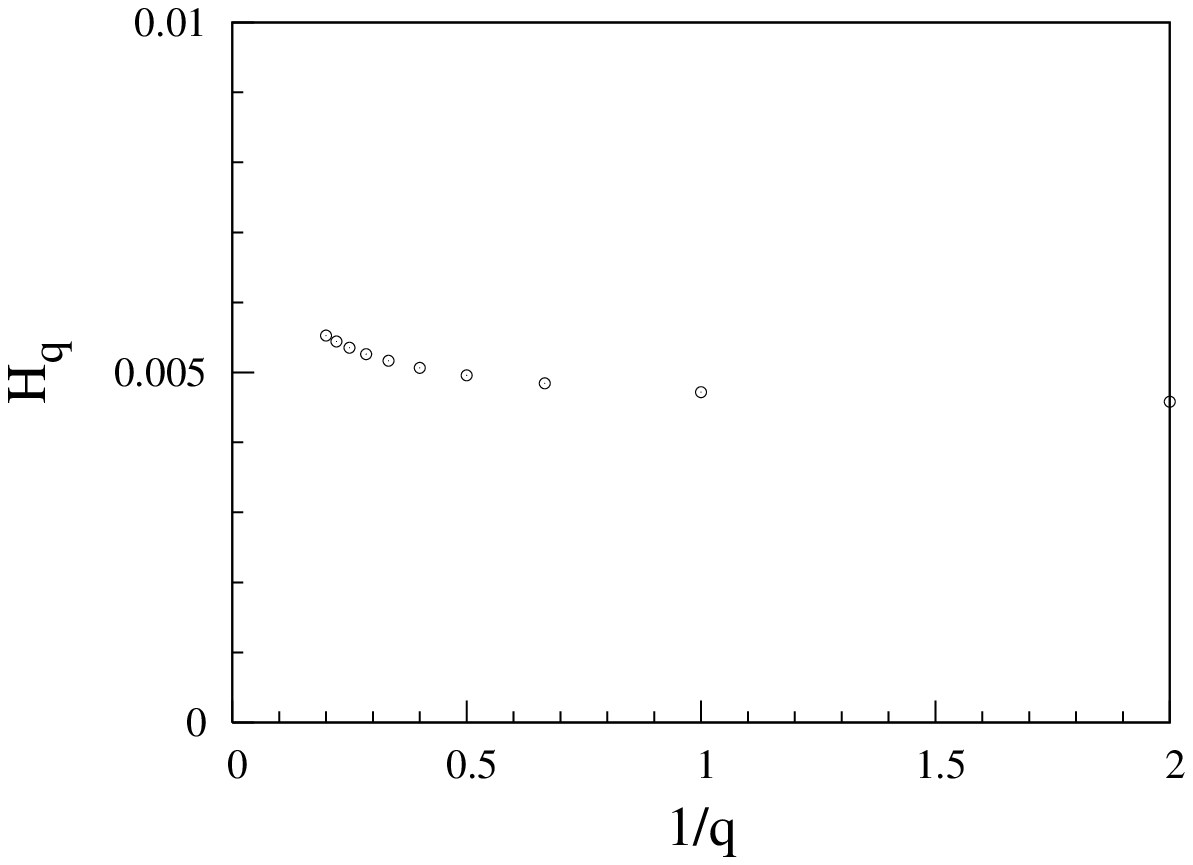}
  \caption{Generalized height-height correlation function $F_{q}(t)$ versus time $t$, (a), and the generalized Hurst 
exponents $H_{q}$ versus $1/q$, (b), of the model \eqref{diff_gamma} and \eqref{diff_x} with $\mu=0$ and $\sigma_{\gamma}=2\cdot 10^{-4}$.}
\label{fig:fract}
\end{figure}

Calculation results shown in Fig. \ref{fig:fract} indicate that the signal is monofractal with the Hurst exponent $H\approx 0$ and the slope of the spectrum $\beta=2H+1$, as for random walk. 

\section{Conclusion}

The linear stochastic differential equation with the slowly fluctuating relaxation may generate monofractal signals with the non-Gaussian $1/f^{\beta}$ noise.


\begin{theacknowledgments}
We acknowledge the support by the Agency for International Science and Technology Development Programs in Lithuania and EU COST Action P10 ``Physics of Risk''.
\end{theacknowledgments}



\bibliographystyle{aipproc}   


\IfFileExists{\jobname.bbl}{}
 {\typeout{}
  \typeout{******************************************}
  \typeout{** Please run "bibtex \jobname" to optain}
  \typeout{** the bibliography and then re-run LaTeX}
  \typeout{** twice to fix the references!}
  \typeout{******************************************}
  \typeout{}
 }



\end{document}